\documentclass[%
 reprint,
 onecolumn, 
 11pt,
nofootinbib,
 amsmath,amssymb,
 aps,
prd,
]{revtex4-2}

\usepackage{graphicx}
\usepackage{dcolumn}
\usepackage{bm}

\usepackage{lineno}
\usepackage{acro}
\usepackage{color}

\DeclareAcronym{DESI}{
  short = DESI ,
  long = Dark Energy Spectroscopic Instrument ,
  short-plural =  ,
}
\DeclareAcronym{DR2}{
  short = DR2 ,
  long = Data Release 2 ,
  short-plural =  ,
}
\DeclareAcronym{CMB}{
  short = CMB ,
  long = cosmic microwave background ,
  short-plural =  ,
}
\DeclareAcronym{BAO}{
  short = BAO ,
  long = baryon acoustic oscillation ,
  short-plural =  ,
}
\DeclareAcronym{BBN}{
  short = BBN ,
  long = Big Bang nucleosynthesis ,
  short-plural =  ,
}

\DeclareAcronym{NH}{
  short = NH ,
  long = normal mass hierarchy ,
  short-plural =  ,
}
\DeclareAcronym{IH}{
  short = IH ,
  long = inverted mass hierarchy ,
  short-plural =  ,
}
\DeclareAcronym{DH}{
  short = DH ,
  long = degenerate mass hierarchy ,
  short-plural =  ,
}
\DeclareAcronym{CL}{
  short = C.L. ,
  long = confidence level ,
  short-plural =  ,
}
\DeclareAcronym{ISW}{
  short = ISW ,
  long = Integrated Sachs–Wolfe ,
  short-plural =  ,
}

\begin{document}


\title{Constraints on the lepton asymmetry from DESI DR2 BAO data}

\author{Zhi-Chao Zhao
}
 \email{zhaozc@cau.edu.cn}
\affiliation{%
Department of Applied Physics, College of Science, China Agricultural University, 17 Qinghua East Road, Haidian District, Beijing 100083, China}%
\author{Dong-Mei Xia}%
\thanks{Corresponding author}%
 \email{xiadm@cqu.edu.cn}
\affiliation{%
Key Laboratory of Low-grade Energy Utilization Technologies and Systems, Ministry of Education, Chongqing University, Chongqing 400044, China}%
\affiliation{%
Department of Nuclear Engineering and Technology, Chongqing University, Chongqing 400044, China}%
\author{Sai Wang}%
 \email{wangsai@hznu.edu.cn}
\affiliation{
School of Physics, Hangzhou Normal University, No.2318 Yuhangtang Road, Yuhang District, Hangzhou 311121, China}

%


\begin{abstract}
It is important to explore the potential existence of lepton asymmetry in the neutrino sector. Conducting a joint analysis of DESI DR2 BAO data and \emph{Planck} 2018 CMB data, we obtain the upper limits on the neutrino degeneracy parameter, i.e., $\xi<0.56$ for the normal mass hierarchy while $\xi<0.62$ for the inverted mass hierarchy, at 95\% confidence level. Considering the influence of the dynamical dark energy, we find that these upper limits remain to be robust. This work may provide helpful implications for model buildings of the matter-antimatter asymmetry in the universe. 
\end{abstract}

\maketitle


\section{Introduction}\label{sec:introduction}

The observed matter-antimatter asymmetry in the universe is a fundamental puzzle in cosmology and particle physics \cite{Canetti:2012zc}. While the baryon asymmetry is relatively well-constrained, the lepton asymmetry remains poorly understood. A significant lepton asymmetry in the early universe could have profound implications for various cosmological phenomena, and importantly, it is considered a potential explanation for the observed matter-antimatter asymmetry. Given the requirement of cosmic charge neutrality, any substantial lepton asymmetry is expected to primarily reside in the neutrino sector \cite{Dolgov:2002wy}.

Cosmological observations, such as the \ac{CMB} and \ac{BAO}, offer valuable tools for constraining lepton asymmetry and the properties of neutrinos \cite{Dolgov:2002wy}. They are sensitive to the energy density of neutrinos and any deviations from the standard model, which could be caused by a non-zero neutrino chemical potential ($\mu_{\nu}$) \cite{Lesgourgues:1999wu}. It is convenient to define neutrino degeneracy parameter $\xi_{\nu}=\mu_{\nu}/T_{\nu}$ with $T_{\nu}$ being temperature. Though the degeneracy parameter of electron-type neutrinos ($\nu_{e}$) has been well constrained ($\xi_{\nu_{e}}\ll 1$) \cite{Kohri:1996ke,Burns:2022hkq,Escudero:2022okz,Domcke:2022uue,Li:2024gzf,Kumar:2022vee}, those of muon-type and tau-type neutrinos ($\nu_{\mu}$ and $\nu_{\tau}$) remain poorly constrained, and they can still be relevantly large, say, $\xi_{\nu_{\mu}}\sim \mathcal{O}(1)$ and $\xi_{\nu_{\tau}}\sim \mathcal{O}(1)$ \cite{Barenboim:2016lxv,Yeung:2020zde,Yeung:2024krv}. While current constraints on the neutrino degeneracy parameter are relatively weak, ongoing efforts are expected to refine these measurements \cite{Kohri:2014hea,SKA-JapanConsortiumCosmologyScienceWorkingGroup:2016hhj,Escudero:2022okz,Bonilla:2018nau}.

Besides analyzing the \ac{CMB} data from \emph{Planck} 2018 results \cite{Planck:2018vyg}, this study analyzes the \ac{BAO} measurements from the \ac{DESI} \ac{DR2} \cite{DESI:2025zgx} to investigate the neutrino degeneracy parameter. \ac{DESI} \ac{DR2} provides the largest dataset ever used for \ac{BAO} measurements, significantly enhancing our ability to probe the expansion history of the universe. While \ac{DESI} \ac{DR2} has already provided new constraints on the sum of neutrino masses ($\sum m_{\nu}$) \cite{DESI:2025ejh} \footnote{Analyzing the \ac{DESI} DR1 data also led to tight constraints on this parameter \cite{Jiang:2024viw}.}, this work specifically focuses on using this new dataset to constrain the neutrino degeneracy parameter. The chemical potential is directly related to the number density asymmetry between neutrinos and antineutrinos. By focusing on this parameter, we could uncover their potential role in addressing the observed matter-antimatter asymmetry in the universe.

The remainder of this paper is structured as follows. In Section \ref{sec:method}, we will demonstrate the cosmological models to investigate and the observational data to analyze. In Section \ref{sec:result}, we will present the results from parameter inference, revealing up-to-date constraints on the neutrino degeneracy. In Section \ref{sec:summary}, we will reveal the main conclusions of this work.

\section{Methodology}\label{sec:method}

\subsection{Theory}

We first demonstrate the properties of neutrinos, i.e., the sum of neutrino masses and the neutrino degeneracy parameter, which are focused on in this work. 
On the one hand, since \ac{DESI} \ac{DR2} placed the tightest upper limits on $\sum m_{\nu}$, in which $m_{\nu}$ denotes the mass of $\nu$ neutrinos, it is necessary for this present work to take into account the mass splittings between neutrino mass eigenstates, which have been indicated by phenomena of neutrino oscillation. 
The measured mass-squared differences are given by \footnote{{https://pdg.lbl.gov}}
\begin{eqnarray}
\Delta m_{21}^{2} &=& \left(+7.49\pm0.19\right)\times 10^{-5} \mathrm{eV}^{2}\,,\label{eq:ms1}\\
\Delta m_{31}^{2} &=& \left(+2.513\pm0.020\right)\times 10^{-3} \mathrm{eV}^{2}\,,\quad(\mathrm{NH})\label{eq:ms2}\\
\Delta m_{32}^{2} &=& \left(-2.484\pm0.020\right)\times 10^{-3} \mathrm{eV}^{2}\,,\quad(\mathrm{IH})\label{eq:ms3}
\end{eqnarray} 
where we introduce \ac{NH} and \ac{IH}. 
{\color{black}On the other hand, since the electron-type neutrino degeneracy is negligible and there is strong mixing between the muon-type and tau-type neutrinos, we assume \cite{Barenboim:2016lxv}
\begin{eqnarray}
\xi_{\nu_{e}}=0\,,\quad \xi_{\nu_{\mu}}=\xi_{\nu_{\tau}}\equiv \xi\,.
\end{eqnarray}
Hence, there is only one independent parameter to characterize the neutrino degeneracy, which is chosen as $\xi$ here and hereafter. 
}

Using the publicly-available \texttt{class} software \cite{Blas:2011rf}, we consider the influence of relic neutrinos on the evolution of the universe. 
The energy density and pressure for a neutrino species are given by
\begin{eqnarray}
\rho_{\nu}+\rho_{\bar{\nu}} &=& \frac{1}{2\pi^{2}}\int_{0}^{\infty}p^{2}dp\sqrt{p^{2}+m_{\nu}^{2}}\left[f_{\nu}(p)+f_{\bar{\nu}}(p)\right] \,,\\
p_{\nu}+p_{\bar{\nu}} &=& \frac{1}{2\pi^{2}}\int_{0}^{\infty}p^{2}dp\frac{p^{2}}{3\sqrt{p^{2}+m_{\nu}^{2}}}\left[f_{\nu}(p)+f_{\bar{\nu}}(p)\right] \,,
\end{eqnarray}
where we introduce $p=|\bold{p}|$ with $\bold{p}$ being a momentum, and $f_{\nu}(p)$ and $f_{\bar{\nu}}(p)$ stand for the distribution functions, i.e., 
\begin{eqnarray}
f_{\nu}(p) &=& \frac{1}{e^{\frac{p}{{T_{\nu}}}+\xi_{\nu}}+1} \,,\\
f_{\bar{\nu}}(p) &=& \frac{1}{e^{\frac{p}{{T_{\nu}}}-\xi_{\nu}}+1} \,.
\end{eqnarray}  
At the conformal time $\tau$ and the spatial position $\bold{x}$, the perturbed distribution function is given by
\begin{equation}
\delta f_{\nu}(\tau,\bold{x},\bold{p}) + \delta f_{\nu}(\tau,\bold{x},\bold{p}) = \left[ \bar{f}_{\nu}(p) + \bar{f}_{\bar{\nu}}(p) \right] \Psi_{\nu}(\tau,\bold{x},\bold{p})\,,
\end{equation}
where $\bar{f}_{\nu}$ and $\bar{f}_{\bar{\nu}}$ stand for the background distribution functions, and $\Psi_{\nu}$ the linear perturbations. 
In the synchronous gauge, the Boltzmann equation for $\Psi_{\nu}$ with Fourier mode $\bold{k}$ is given by 
\begin{equation}\label{eq:10}
\dot{\Psi}_{\nu} + i\frac{y}{y^{2}+a^{2}\tilde{m}_{\nu}^{2}} \left(\bold{k}\cdot\hat{\bold{n}}\right) \Psi_{\nu} + \frac{d\ln\left(\bar{f}_{\nu}+\bar{f}_{\bar{\nu}}\right)}{d\ln y} \left[\dot{\eta}_{\mathrm{T}} - \frac{1}{2}\left(\dot{h}_{\mathrm{L}}+6\dot{\eta}_{\mathrm{L}}\right)\left(\bold{k}\cdot\hat{\bold{n}}\right)^{2}\right] = 0 \,,
\end{equation}
where $\eta_{\mathrm{T}}$ and $h_{\mathrm{L}}$ stand for the metric perturbations, an overdot the derivative with respect to $\tau$, and $\hat{\bold{n}}=\bold{p}/p$ the direction of momentum. For simplicity, we have introduced $y=ap/T_{\nu0}$ and $\tilde{m}_{\nu}=m_{\nu}/T_{\nu0}$ with $a$ being the scale factor of the universe and $T_{\nu0}=aT_{\nu}$ the present-day temperature.  
The neutrino chemical potential contributes to the Boltzmann equation through the following factor
\begin{equation}\label{eq:11}
\frac{d\ln\left(\bar{f}_{\nu}+\bar{f}_{\bar{\nu}}\right)}{d\ln y} = - \frac{y\left(1+\cosh\xi\cosh y\right)}{\left(\cosh\xi+e^{-y}\right)\left(\cosh\xi+\cosh y\right)}\ .
\end{equation}
For readers who are interested in details of this topic, we encourage them to refer to Refs.~\cite{Lesgourgues:1999wu,Kohri:2014hea}.

{\color{black}

Neutrino degeneracy changes not only the radiation energy density of the Universe, but also the evolution of linear cosmological perturbations. For massless neutrinos, its effect is primarily through the background expansion and can be fully captured by an excess in the effective number of relativistic species $N_{\rm eff}$, i.e., $\Delta N_{\rm eff}$. For massive neutrinos, however, neutrino degeneracy impacts both the background and the perturbations. Through Eq.~(\ref{eq:11}), it modifies the phase–space distribution entering the Boltzmann hierarchy in Eq.~(\ref{eq:10}). Therefore, a full numerical solution is required, as implemented in this work. The above effects propagate to the radiation and matter power spectra and can thus be constrained with \ac{CMB} and \ac{BAO} data. If neutrinos have only tiny masses, the resulting constraints on $\xi$ approach those inferred from $\Delta N_{\rm eff}$ alone. For any finite mass, the degeneracy with $\Delta N_{\rm eff}$ is broken to some extent, so bounds on $\xi$ cannot be read off from $\Delta N_{\rm eff}$ alone and should be obtained with the full Boltzmann treatment. 
}

Since \ac{DESI} \ac{DR2} \ac{BAO} data also suggested strong evidence for the dynamical dark energy \cite{DESI:2025fii,Gu:2025xie}, we take it into account in \texttt{class} via using the CPL parameterization \cite{Chevallier:2000qy,Linder:2002et}, i.e., the equation of state 
\begin{equation}\label{eq:eos}
w(z)=w_{0}+w_{a}\frac{z}{1+z}\,,
\end{equation}
with $z$ being the cosmological redshift while $w_{0}$ and $w_{a}$ two independent parameters to infer in the following sections. When choosing $w_{0}=-1$ and $w_{a}=0$, we recover the cosmological constant ($\Lambda$).

In summary, we mainly focus on two classes of cosmological models. The first one refers to $\Lambda$CDM plus massive neutrinos with chemical potential, which is denoted as ``$\Lambda\mathrm{CDM}$+$\sum m_{\nu}$+$\xi$'' hereafter. They have eight independent parameters, i.e., 
\begin{equation}
    \Big\{\Omega_{b}h^{2}, \Omega_{c}h^{2}, 100\theta_{\mathrm{MC}}, \ln(10^{10}A_{s}), n_{s}, \tau_{\mathrm{reio}}, \sum m_{\nu}, \xi\Big\}\,. 
\end{equation}
Here, the present-day physical energy-density fractions of baryons and cold dark matter are denoted as $\Omega_{b}h^{2}$ and $\Omega_{c}h^{2}$, respectively. The ratio between the sound horizon and the angular diameter distance at the decoupling epoch is $\theta_{\mathrm{MC}}$. The Thomson scatter optical depth due to reionization is $\tau_{\mathrm{reio}}$. The amplitude and index of the power spectrum of primordial curvature perturbations at the pivot scale $k_{p}=0.05\,\mathrm{Mpc}^{-1}$ are denoted as $A_{s}$ and $n_{s}$, respectively. The second class of models refer to $\mathrm{CPL}$ plus massive neutrinos with chemical potential, which is denoted as ``$\mathrm{CPL}$+$\sum m_{\nu}$+$\xi$'' hereafter. Besides the aforementioned eight independent parameters, they have two additional independent parameters, i.e., 
\begin{equation}
\Big\{w_{0}, w_{a}\Big\}\,,
\end{equation}
which are defined in Eq.~(\ref{eq:eos}). As suggested in Refs.~\cite{Huang:2015wrx,Wang:2016tsz}, it should be also noted that the neutrino mass hierarchies lead to lower bounds on the priors of $\sum m_{\nu}$, i.e., 
\begin{eqnarray}
&&\sum m_{\nu}>0.05878\,\mathrm{eV}\,,\quad(\mathrm{NH})\\
&&\sum m_{\nu}>0.09892\,\mathrm{eV}\,.\quad(\mathrm{IH})
\end{eqnarray} 
For simplicity, we have neglected the uncertainties of neutrino mass-squared differences, as shown in Eqs.~(\ref{eq:ms1},\ref{eq:ms2},\ref{eq:ms3}), on the above lower bounds, since they are expected not to change our results. 
{\color{black}In addition, to check whether the constraints on $\xi$ strongly depend on the assumption of NH/IH, we further consider the \ac{DH}, i.e., $m_{1}=m_{2}=m_{3}$. In this case, we use the prior as follows 
\begin{equation}
    \sum m_{\nu}>0\,\mathrm{eV}\,.\quad(\mathrm{DH})
\end{equation}
Furthermore, we consider a third class of models that involves $\Delta N_{\rm eff}$ as an additional independent parameter, since $\xi$ is expected to contribute to $\Delta N_{\rm eff}$, as mentioned above. In the following, we denote these models as ``$\Lambda\mathrm{CDM}$+$\sum m_{\nu}$+$\xi$+$\Delta N_{\rm eff}$'' and ``$\mathrm{CPL}$+$\sum m_{\nu}$+$\xi$+$\Delta N_{\rm eff}$'', respectively. 
}

\subsection{Data analysis}

We then fit the above cosmological models to a combination of the cutting-edge \ac{CMB} and \ac{BAO} data. We utilize the \emph{Planck} 2018 results of \ac{CMB} temperature anisotropies and polarization \cite{Planck:2019nip}. To be specific, we adopt the \texttt{plik} likelihoods of high-$\ell$ temperature and polarization auto and cross angular power spectra, denoted as ``high-$\ell$ TTTEEE'', as well as the \texttt{SimAll} and \texttt{Commander} likelihoods of low-$\ell$ temperature and polarization auto angular power spectrum, as denoted as ``low-$\ell$ TTEE''. We further utilize the up-to-date \ac{BAO} observational data from \ac{DESI} \ac{DR2}, which has integrated the largest dataset of galaxy and spectroscopic redshift measurements, and Lyman-$\alpha$ forest spectra, as well as their cross-correlation with quasar positions \cite{DESI:2025zgx}. 
Finally, we utilize the \texttt{MontePython} software \cite{Brinckmann:2018cvx,Audren:2012wb} to infer the cosmological parameters.

\section{Results}\label{sec:result}

\begin{figure*}
    \centering
    \includegraphics[width=0.9\linewidth]{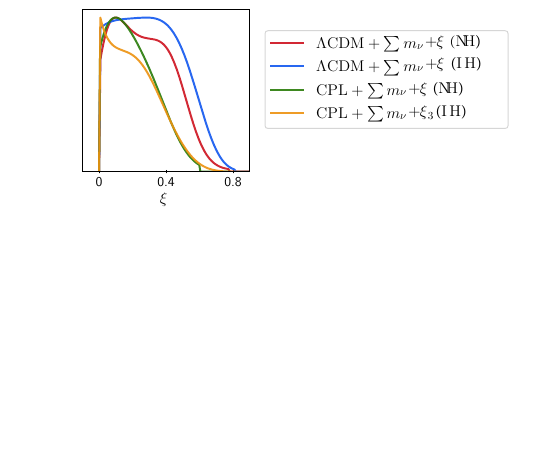}
    \caption{One-dimensional posterior distributions of \(\xi\). Results are shown for \(\Lambda\mathrm{CDM}\)\(+\Sigma m_\nu + \xi\) and CPL\(+\Sigma m_\nu + \xi\), each assuming either NH or IH. 
}
    \label{fig:1d}
\end{figure*}

\begin{figure*}
    \centering
    \includegraphics[width=0.9\linewidth]{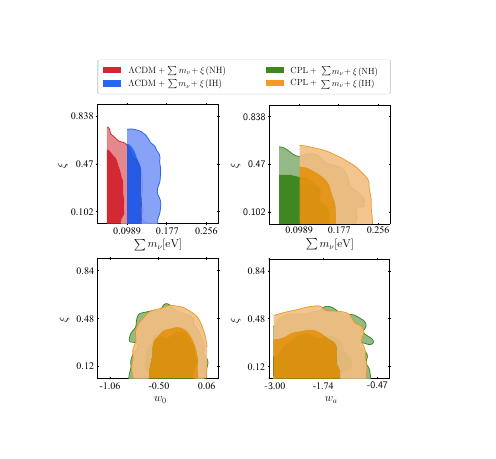}
    \caption{Two-dimensional posterior distributions for $\xi$ versus $\sum m_\nu$ (upper panels), $w_0$ (bottom left panel), and $w_a$ (bottom right panel). Light and dark shaded regions, respectively, stand for 68\% C.L. and 95\% C.L.. Results are shown for \(\Lambda\mathrm{CDM}\)\(+\Sigma m_\nu + \xi\) and CPL\(+\Sigma m_\nu + \xi\), each assuming either NH or IH. 
}
    \label{fig:2d}
\end{figure*}

The results from the parameter inference are shown in Tables~\ref{tab:results}, \ref{tab:results_Nur}, and Figures~\ref{fig:1d}, \ref{fig:2d}. To be specific, we summarize the mean value and uncertainties of all the independent parameters as well as the minimal $\chi^{2}$ in Tables~\ref{tab:results} and \ref{tab:results_Nur}. Throughout this work, all the uncertainties are shown at 68\% \ac{CL} while all the upper limits are shown at 95\% \ac{CL}. In Figure~\ref{fig:1d}, we exhibit the likelihood distributions of $\xi$ for the two classes of cosmological models inferred in this work. In Figure~\ref{fig:2d}, we exhibit the contours in two-dimensional planes spanned by $\xi$ and either $\sum m_{{\nu}}$, or $w_{0}$, or $w_{a}$. They are depicted as the dark shaded regions for 68\% \ac{CL}, while the light shaded regions for 95\% \ac{CL}.

\begin{table*}[htbp]
  \centering
\begin{tabular}{|l|l|l|l|l|l|l|}
\hline
& \multicolumn{3}{c|}{$\Lambda\mathrm{CDM} + \sum m_{\nu} + \xi$}
& \multicolumn{3}{c|}{$\mathrm{CPL} + \sum m_{\nu} + \xi$} \\ 
\cline{2-7}
  & NH & IH & DH & NH & IH & DH \\ 
\hline 
  $100 ~ \Omega_b h^2$ & $2.260_{- 0.015}^{+ 0.014}$ & $2.265_{- 0.016}^{+
  0.015}$ & $2.258_{- 0.016}^{+ 0.015}$ & $2.243_{- 0.014}^{+ 0.014}$ &
  $2.244_{- 0.015}^{+ 0.014}$ & $2.242_{- 0.016}^{+ 0.015}$\\
  $\Omega_c h^2$ & $0.1188_{- 0.0021}^{+ 0.0009}$ & $0.1187_{- 0.0027}^{+
  0.0012}$ & $0.1194_{- 0.0022}^{+ 0.0013}$ & $0.1204_{- 0.0016}^{+ 0.0011}$ &
  $0.1204_{- 0.0016}^{+ 0.0011}$ & $0.1208_{- 0.0017}^{+ 0.0013}$\\
  $100 \theta_{\rm{MC}}$ & $1.042_{- 0.00030}^{+ 0.00038}$ & $1.042_{-
  0.00035}^{+ 0.00044}$ & $1.042_{- 0.00034}^{+ 0.00041}$ & $1.042_{-
  0.00030}^{+ 0.00034}$ & $1.042_{- 0.00031}^{+ 0.00033}$ & $1.042_{-
  0.00031}^{+ 0.00036}$\\
  $\ln (10^{10} A_s)$ & $3.051_{- 0.019}^{+ 0.016}$ & $3.053_{- 0.021}^{+
  0.019}$ & $3.051_{- 0.021}^{+ 0.018}$ & $3.047_{- 0.016}^{+ 0.015}$ &
  $3.048_{- 0.018}^{+ 0.016}$ & $3.047_{- 0.017}^{+ 0.017}$\\
  $n_s$ & $0.9745_{- 0.0047}^{+ 0.0040}$ & $0.9762_{- 0.0058}^{+ 0.0045}$ &
  $0.9730_{- 0.0052}^{+ 0.0044}$ & $0.9679_{- 0.0046}^{+ 0.0040}$ & $0.9686_{-
  0.0048}^{+ 0.0041}$ & $0.9676_{- 0.0051}^{+ 0.0043}$\\
  $\tau_{\rm{reio}}$ & $0.0588_{- 0.0090}^{+ 0.0071}$ & $0.0598_{-
  0.0093}^{+ 0.0096}$ & $0.0579_{- 0.0097}^{+ 0.0082}$ & $0.0548_{- 0.0078}^{+
  0.0072}$ & $0.0555_{- 0.0082}^{+ 0.0077}$ & $0.0545_{- 0.0084}^{+ 0.0078}$\\
  $w_0$ & $\slash$ & $\slash$ & $\slash$ & $- 0.37_{- 0.19}^{+ 0.23}$ & $-
  0.34_{- 0.16}^{+ 0.24}$ & $- 0.43_{- 0.22}^{+ 0.25}$\\
  $w_a$ & $\slash$ & $\slash$ & $\slash$ & $- 1.91_{- 0.72}^{+ 0.54}$ & $-
  2.04_{- 0.79}^{+ 0.41}$ & $- 1.72_{- 0.73}^{+ 0.67}$\\
  $\sum m_{\nu} \hspace{0.17em}$[eV] & $< 0.120$ & $< 0.157$ & $< 0.090$ & $<
  0.196$ & $< 0.218$ & $< 0.173$\\
  $\xi$ & $< 0.56$ & $< 0.62$ & $< 0.58$ & $< 0.47$ & $< 0.48$ & $< 0.49$\\
  \hline
  $\chi^2_{\min}$ & $2787$ & $2791$ & 2785 & $2777$ & $2779$ & 2777\\
  \hline
\end{tabular}
\caption{Constraints on the cosmological parameters. We show the uncertainties of model parameters at 68\% C.L., while the upper limits at 95\% C.L.. The minimal $\chi^{2}$ is shown for model comparison. }  \label{tab:results}
\end{table*}

\begin{table*}[htbp]
  \centering
\begin{tabular}{|l|l|l|l|l|l|l|}
\hline
& \multicolumn{3}{c|}{$\Lambda\mathrm{CDM} + \sum m_{\nu} + \xi + \Delta N_{\rm{eff}}$}
& \multicolumn{3}{c|}{$\mathrm{CPL} + \sum m_{\nu} + \xi + \Delta N_{\rm{eff}}$} \\ 
\cline{2-7}
  & NH & IH & DH & NH & IH & DH \\ 
\hline 
  $100 ~ \Omega_b h^2$ & $2.268_{- 0.017}^{+ 0.015}$ & $2.272_{- 0.016}^{+
  0.016}$ & $2.264_{- 0.015}^{+ 0.015}$ & $2.25_{- 0.016}^{+ 0.015}$ &
  $2.25_{- 0.017}^{+ 0.015}$ & $2.248_{- 0.016}^{+ 0.016}$\\
  $\Omega_c h^2$ & $0.1208_{- 0.0028}^{+ 0.0017}$ & $0.1209_{- 0.0029}^{+
  0.0020}$ & $0.1211_{- 0.0026}^{+ 0.0016}$ & $0.1220_{- 0.0022}^{+ 0.0015}$ &
  $0.1219_{- 0.0023}^{+ 0.0015}$ & $0.1221_{- 0.0021}^{+ 0.0016}$\\
  $100 \theta_{\rm{MC}}$ & $1.042_{- 0.00040}^{+ 0.00044}$ & $1.042_{-
  0.00039}^{+ 0.00044}$ & $1.042_{- 0.00039}^{+ 0.00042}$ & $1.042_{-
  0.00033}^{+ 0.00037}$ & $1.042_{- 0.00035}^{+ 0.00039}$ & $1.042_{-
  0.00034}^{+ 0.00038}$\\
  $\ln (10^{10} A_s)$ & $3.057_{- 0.019}^{+ 0.017}$ & $3.059_{- 0.020}^{+
  0.017}$ & $3.056_{- 0.019}^{+ 0.018}$ & $3.052_{- 0.017}^{+ 0.016}$ &
  $3.053_{- 0.018}^{+ 0.017}$ & $3.051_{- 0.018}^{+ 0.017}$\\
  $n_s$ & $0.9780_{- 0.0060}^{+ 0.0044}$ & $0.9798_{- 0.0058}^{+ 0.0049}$ &
  $0.9761_{- 0.0054}^{+ 0.0046}$ & $0.971_{- 0.0058}^{+ 0.0047}$ & $0.9716_{-
  0.0059}^{+ 0.0047}$ & $0.9705_{- 0.0059}^{+ 0.0045}$\\
  $\tau_{\rm{reio}}$ & $0.0595_{- 0.0093}^{+ 0.0075}$ & $0.0603_{-
  0.0089}^{+ 0.0080}$ & $0.0582_{- 0.0089}^{+ 0.0081}$ & $0.0557_{- 0.0084}^{+
  0.0076}$ & $0.0561_{- 0.0081}^{+ 0.0077}$ & $0.0550_{- 0.0085}^{+ 0.0079}$\\
  $w_0$ & $\slash$ & $\slash$ & $\slash$ & $- 0.39_{- 0.19}^{+ 0.25}$ & $-
  0.37_{- 0.18}^{+ 0.24}$ & $- 0.45_{- 0.23}^{+ 0.23}$\\
  $w_a$ & $\slash$ & $\slash$ & $\slash$ & $- 1.85_{- 0.72}^{+ 0.63}$ & $-
  1.93_{- 0.72}^{+ 0.56}$ & $- 1.65_{- 0.69}^{+ 0.71}$\\
  $\sum m_{\nu} \hspace{0.17em}$[eV] & $< 0.134$ & $< 0.164$ & $< 0.0889$ & $<
  0.215$ & $< 0.223$ & $< 0.182$\\
  $\xi$ & $< 0.53$ & $< 0.54$ & $< 0.48$ & $< 0.45$ & $< 0.45$ & $< 0.44$\\
  \hline
  $\Delta N_{\rm{eff}}$ & $< 0.37$ & $< 0.41$ & $< 0.35$ & $< 0.28$ & $< 0.30$ & $<
  0.27$\\
  \hline
  $\chi^2_{\min}$ & $2789$ & 2791 & $2784$ & 2780 & 2781 & $2778$\\
  \hline
\end{tabular}
\caption{\color{black}Same with Tab.~\ref{tab:results}, but including $\Delta N_{\rm eff}$ as an additional independent parameter in the analysis. }  \label{tab:results_Nur}
\end{table*}

For the $\Lambda\mathrm{CDM}$+$\sum m_{\nu}$+$\xi$ model, we find the upper limits on the neutrino degeneracy parameter as follows 
\begin{eqnarray}
&&\xi < 0.56 \quad (95\%\,\mathrm{C.L.};\ \mathrm{NH})\ , \\
&&\xi < 0.62 \quad (95\%\,\mathrm{C.L.};\ \mathrm{IH})\ , \\
&&\xi < 0.58 \quad (95\%\,\mathrm{C.L.};\ \mathrm{DH})\ .
\end{eqnarray}
The upper limits on $\sum m_{\nu}$ inferred in this present work are compatible with those inferred by \ac{DESI} \ac{DR2} \cite{DESI:2025ejh}, which missed the neutrino degeneracy in consideration.

For the $\mathrm{CPL}$+$\sum m_{\nu}$+$\xi$ model, we find the upper limits on the neutrino degeneracy parameter as follows 
\begin{eqnarray}
&&\xi < 0.47 \quad (95\%\,\mathrm{C.L.};\ \mathrm{NH})\ , \\
&&\xi < 0.48 \quad (95\%\,\mathrm{C.L.};\ \mathrm{IH})\ , \\
&&\xi < 0.49 \quad (95\%\,\mathrm{C.L.};\ \mathrm{DH})\ . 
\end{eqnarray}
The upper limits on $\sum m_{\nu}$ and the constraints on $w_{0}$ and $w_{a}$ inferred in this present work are also compatible with those inferred by \ac{DESI} \ac{DR2} \cite{DESI:2025ejh}, which again missed the neutrino degeneracy in consideration.

Through comparing the above results with each other, we find that taking the dynamical dark energy into consideration can slightly tighten the upper limits on $\xi$. This conclusion is still compatible with our expectation. It could be roughly explained by correlations between $\xi$ and other parameters, i.e., $w_{0}$ and $w_{a}$, as revealed in Figure~\ref{fig:2d}. In addition, based on the differences between the minimal $\chi^{2}$, the second class of models seem to fit the observational data slightly better than the first class of models. However, it should be noted that the former have two more parameters than the latter. 
{\color{black}Furthermore, comparing the results for DH with those for either NH or IH, we find that the constraints on $\xi$ do not strongly depend on the neutrino mass hierarchy. }

\begin{figure}[t]
\centering
\includegraphics[width=0.9\linewidth]{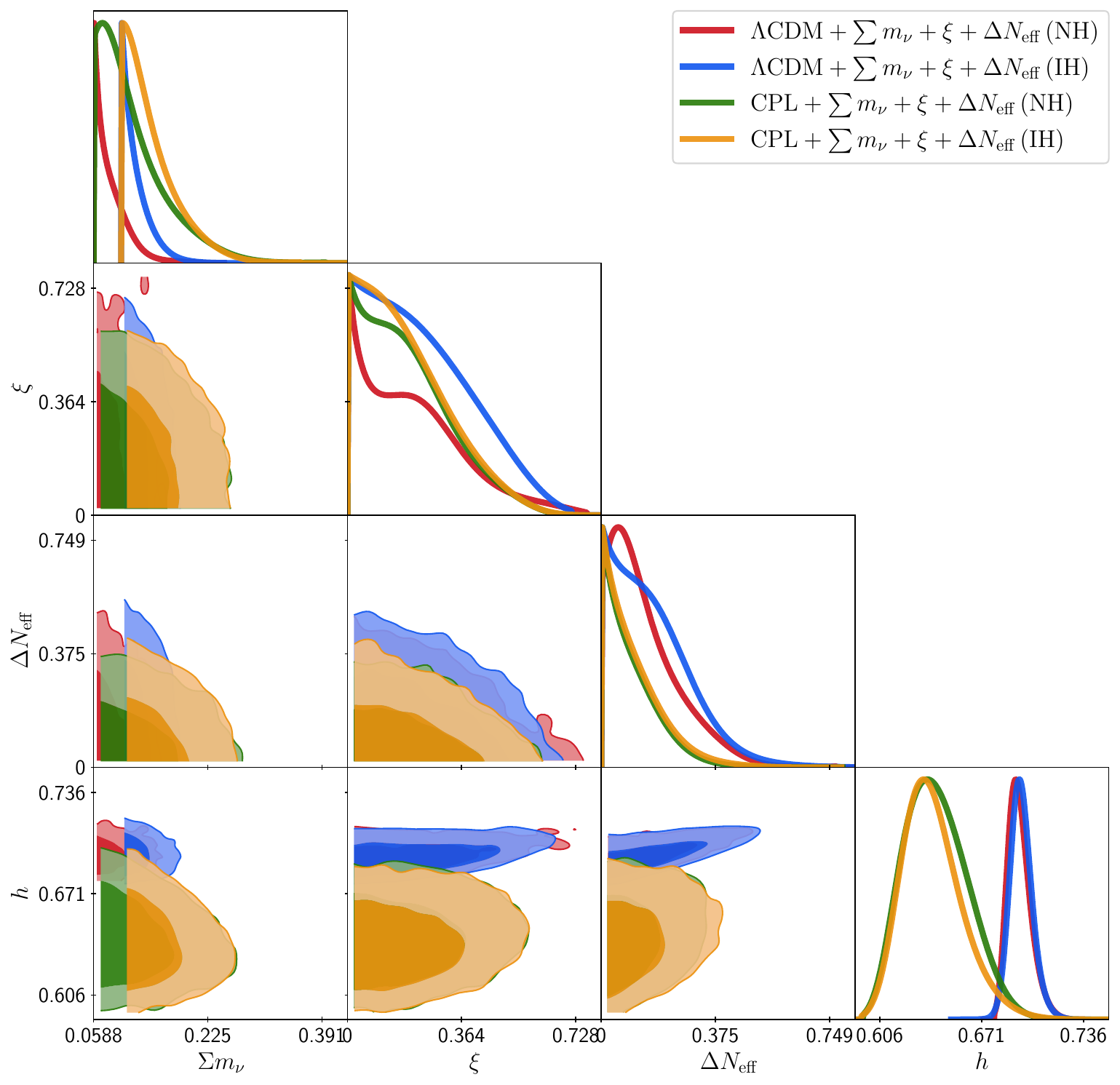}
\caption{\color{black}Triangle plot for the parameters $\sum m_\nu,\xi,\Delta N_{\rm eff},{\rm and~ }h$. Light and dark shaded regions, respectively, stand for 68\% C.L. and 95\% C.L.. Results are shown for$\Lambda$CDM$+\sum m_\nu+\xi+\Delta N_{\rm eff}$ and CPL$+\sum m_\nu+\xi+\Delta N_{\rm eff}$, each assuming either NH or IH.}
\label{fig:replyref1}
\end{figure}

{\color{black}
The constraints obtained from including $\Delta N_{\rm eff}$ as an additional independent parameter are summarized in Table~\ref{tab:results_Nur}.
Comparing Table~\ref{tab:results_Nur} with Table~\ref{tab:results}, we find slightly stronger (by $\sim10\%$) constraints on $\xi$ when introducing $\Delta N_{\rm eff}$ as a free parameter. To demonstrate possible correlations between $\xi$ and $\Delta N_{\rm eff}$, we further depict a triangle plot for the parameters $\sum m_{\nu}$, $\xi$, $\Delta N_{\rm eff}$, and $h$ in Figure~\ref{fig:replyref1}. 
We find that there is no significant correlation between $\xi$ and $\Delta N_{\rm eff}$, as the two-dimensional contours are nearly aligned with the coordinate axes. }

\section{Summary}\label{sec:summary}

In this work, we obtained the up-to-date upper limits on the neutrino degeneracy parameter via analyzing the data combination of \emph{Planck} 2018 \ac{CMB} and \ac{DESI} \ac{DR2} \ac{BAO}. Assuming the model of $\Lambda$CDM plus massive neutrinos with chemical potential, we found the upper limits to be $\xi<0.56$ (\ac{NH}), $\xi<0.62$ (\ac{IH}), and $\xi<0.58$ (\ac{DH}) at 95\% \ac{CL}. Considering the influence of the dynamical dark energy, we found the upper limits to be $\xi<0.47$ (\ac{NH}), $\xi<0.48$ (\ac{IH}), and $\xi<0.49$ (\ac{DH}), which both become slightly tighter. Comparing the above two sets of results, we revealed that the upper limits on $\xi$ remain to be robust. Our present work might provide helpful implications for model buildings of the matter-antimatter asymmetry in the universe. Future observations of the \ac{CMB} B-mode polarization and the 21 cm fluctuations were expected to significantly improve the measurement precision of the neutrino degeneracy, possibly leading the lepton asymmetry to explain the observed matter-antimatter asymmetry. 

{\color{black}To assess DESI’s contribution to the constraints on $\xi$, we also perform Bayesian parameter inferences without DESI and then compare the derived results with those obtained with DESI obtained in this work. In $\Lambda$CDM, the upper limits on $\xi$ without DESI (with DESI) are given by $0.46$ ($0.56$) for NH, $0.47$ ($0.62$) for IH, and $0.49$ ($0.58$) for DH, indicating weaker constraints when including DESI DR2 data. In contrast, in the CPL model, the limits are $0.51$ ($0.47$) for NH, $0.48$ ($0.48$) for IH, and $0.48$ ($0.49$) for DH. A plausible interpretation is that the DESI DR2 BAO measurements favor late-time deviations from the cosmological constant, indicating dynamical evolution of dark energy. Combining the DESI and \emph{Planck} data within $\Lambda$CDM shifts the posteriors and weakens the constraints on $\xi$, whereas allowing for a two-parameter CPL model of  $w(z)$ largely reconciles the datasets. }

{\color{black}Leaving $\xi_\mu$ and $\xi_\tau$ to vary independently in the Bayesian parameter inferences, we can further find that the upper limits on $\xi$ are weakened by $\simeq 30\%$ (with $\pm 5\%$ variation across hierarchies and between $\Lambda$CDM/CPL). This is expected because CMB+BAO are chiefly sensitive to the even–power combination, e.g., $\xi_\mu^2+\xi_\tau^2$ \cite{Kumar:2022vee}, and introducing an extra nearly-degenerate parameter may inflate the one-dimensional posteriors. The prior $\xi_\mu=\xi_\tau$ is physically motivated by the flavor equilibrium in the early universe \cite{Barenboim:2016lxv}. Relaxing it would not alter one of our qualitative conclusions, that is, the bounds remain robust at the $\sim30\%$ level.
}

\acknowledgments

Z.C.Z. is supported by the National Key Research and Development Program of China Grant No. 2021YFC2203001. S.W. is supported by the National Natural Science Foundation of China (Grant No. 12175243) and the National Key R\&D Program of China No. 2023YFC2206403.


\end{document}